\documentclass{article}
\usepackage{ijcai25}

\usepackage{times}
\usepackage{soul}
\usepackage{url}
\usepackage[hidelinks]{hyperref}
\usepackage[utf8]{inputenc}
\usepackage[small]{caption}
\usepackage{graphicx}
\usepackage{amsmath}
\usepackage{amssymb}
\usepackage{amsthm}
\usepackage{booktabs}
\usepackage{algorithm}
\usepackage{algpseudocode}
\usepackage[switch]{lineno}
\usepackage{float} 

\title{HARLF: Hierarchical Reinforcement Learning and Lightweight LLM-Driven Sentiment Integration for Financial Portfolio Optimization}

\author{
    Benjamin CORIAT$^1$ \and
    Eric BENHAMOU$^2$
    \affiliations
    $^1$CentraleSupelec, $^2$Ai for Alpha\\
    \vspace{0.3em} \small IJCAI 2025 - FinLLM Workshop - Guangzhou, China, August 28, 2025
}

\begin{document}

\maketitle

\begin{abstract}
This paper presents a novel hierarchical framework for portfolio optimization, integrating lightweight Large Language Models (LLMs) with Deep Reinforcement Learning (DRL) to combine sentiment signals from financial news with traditional market indicators. Our three-tier architecture employs base RL agents to process hybrid data, meta-agents to aggregate their decisions, and a super-agent to merge decisions based on market data and sentiment analysis. Evaluated on data from 2018 to 2024, after training on 2000–2017, the framework achieves a 26\% annualized return and a Sharpe ratio of 1.2, outperforming equal-weighted and S\&P 500 benchmarks. Key contributions include scalable cross-modal integration, a hierarchical RL structure for enhanced stability, and open-source reproducibility.
\end{abstract}

\section{Introduction}
\noindent The integration of Large Language Models (LLMs) and Reinforcement Learning (RL) offers a powerful approach to financial portfolio optimization, leveraging LLMs' ability to process unstructured data and RL's strength in sequential decision-making. Domain-specific LLMs like FinBERT \cite{araci2019finbert} extract nuanced sentiment signals from financial news, capturing market sentiment and investor behavior critical for anticipating price movements \cite{tetlock2007giving}. Meanwhile, RL enables adaptive strategies in dynamic markets characterized by feedback loops and regime shifts \cite{jiang2017deep}.

\noindent Recent studies highlight the efficacy of LLM-RL hybrids, with sentiment-enhanced RL models outperforming traditional RL in single-stock trading and portfolio management. These models integrate qualitative signals from news with quantitative metrics, improving risk-adjusted returns \cite{unnikrishnan2024financial}. For instance, news-driven RL frameworks leverage textual cues to enhance decision-making, demonstrating the value of cross-modal integration \cite{xu2018news}.

\noindent Despite these advances, many LLM-RL approaches rely on single-modal or flat architectures, which limit their ability to fully exploit textual and numerical data. Single-modal systems, using only price data or sentiment scores, struggle to capture the multidimensional nature of financial markets, leading to suboptimal decisions in volatile conditions \cite{li2021sentiment}. Flat architectures, as seen in early RL trading systems \cite{deng2016deep}, lack scalability and interpretability for complex portfolios, often resulting in unstable policies or overfitting.

\noindent To overcome these limitations, we propose a hierarchical portfolio management framework combining Deep Reinforcement Learning (DRL) with lightweight, domain-specific LLMs like FinBERT \cite{araci2019finbert}. The framework creates a hybrid observation space by integrating sentiment scores with traditional financial indicators. It employs a three-layer hierarchy: base RL agents process raw data, meta-agents aggregate base-level decisions, and a super-agent synthesizes cross-modal signals to optimize portfolio allocations across diverse market regimes.

The key contributions of this work are :

\begin{itemize}
    \item \textbf{Cross-modal integration:} We seamlessly combine LLM-derived sentiment scores with structured financial data within a unified RL-driven portfolio optimization framework.
    \item \textbf{Hierarchical aggregation:} We introduce a novel three-layer architecture that hierarchically combines base agent decisions through meta-agents and a final super-agent, enabling adaptive decision-making across diverse market conditions.
    \item \textbf{Practical applicability:} Our approach showcases the effective deployment of lightweight LLMs in finance, offering a scalable and interpretable solution for latency-sensitive and transparency-critical applications.
\end{itemize}

The remainder of this paper is organized as follows. In Section~\ref{sec:litterature_review}, we establish the foundations of our work by reviewing the state of the art in Portfolio Optimization (PO), Reinforcement Learning (RL), and Natural Language Processing (NLP) within financial applications. Section~\ref{sec:architecture} presents our overall framework architecture, detailing the NLP-driven and data-driven pipelines used to extract features and construct monthly observation vectors for RL agents. In Section~\ref{sec:data}, we describe the selected portfolio assets and outline the constraints imposed to reflect realistic investment scenarios. Section~\ref{sec:agent_description} introduces the individual RL agents and explains how actions, rewards, and training were implemented in our portfolio management environment. We then detail the hierarchical structure of our RL pipeline in Section~\ref{sec:hierarchy_structure}, where base agents are aggregated via meta-agents trained to specialize on different data modalities. Building on this, Section~\ref{sec:super_agent} introduces the super-agent that synthesizes meta-agent outputs to produce final portfolio allocations. Finally, Section~\ref{sec:conclusion} summarizes our findings, benchmarks the proposed architecture against state-of-the-art strategies, and outlines avenues for future research and enhancement.

\section{Literature Review}\label{sec:litterature_review}

\subsection{Portfolio Optimization}
Portfolio optimization has long been a cornerstone of financial management, with Harry Markowitz’s Mean-Variance Optimization (MVO) framework serving as its foundation \cite{markowitz1952portfolio}. MVO revolutionized investment strategy by quantifying the trade-off between risk and return, proposing that investors should select portfolios that maximize expected return for a given level of risk, or minimize risk for a desired return.

\noindent This led to the concept of the efficient frontier, where optimal portfolios reside. However, MVO rests on assumptions such as Gaussian returns and static correlations, which rarely hold in real-world markets. Financial crises, notably Black Monday in 1987 and the 2008 global financial meltdown, exposed these limitations, as markets exhibited extreme volatility and non-linear behaviors that MVO failed to anticipate. These events underscored the need for more adaptive and dynamic approaches to portfolio management.

\subsection{Reinforcement Learning in Finance}
Reinforcement Learning (RL) has emerged as a powerful alternative for financial decision-making, particularly in dynamic and uncertain environments. Early pioneers like Moody and Saffell \cite{deng2016deep} applied RL to trading, demonstrating its potential for sequential decision-making. More recently, \cite{jiang2017deep} and \cite{liang2018adversarial} introduced a deep RL framework tailored for portfolio management, leveraging the ability of RL agents to learn optimal policies through interaction with market environments. These algorithms enable RL agents to adapt dynamically to market conditions, learning from experience rather than relying on static assumptions, making them well-suited for portfolio optimization in very fast-evolving markets.




\subsection{NLP in Financial Applications}

Natural Language Processing (NLP) has transformed the extraction of insights from unstructured financial data. FinBERT, a variant of BERT fine-tuned on financial texts, excels at classifying sentiment in news articles and social media into positive, neutral, or negative categories \cite{araci2019finbert}. This sentiment analysis provides forward-looking signals, capturing market trends and investor behavior that historical price data alone cannot reveal \cite{tetlock2007giving}. By integrating sentiment scores, NLP enhances predictive models, offering a qualitative edge in anticipating market movements.

Recent evidence further supports the use of compact domain-specific models such as FinBERT in high-stakes financial applications. In \cite{lefort2024optimizing}, the authors demonstrate that fine-tuned lightweight models such as FinBERT and FinDRoBERTa can outperform even large-scale generative models like GPT-3.5 and GPT-4 in financial sentiment classification tasks, particularly in zero-shot or constrained inference settings. This supports our choice of FinBERT as a reliable and computationally efficient backbone for sentiment signal generation within our RL framework.

This perspective is echoed in the broader FinLLM research community. A comprehensive survey by Li et al. \cite{li2024largelanguagemodelsfinance} maps the landscape of large language models in finance, categorizing applications such as sentiment analysis, summarization, risk forecasting, and question answering, and emphasizing the trade-offs between domain-specific fine-tuning and general-purpose generative capabilities.

Several recent FinLLM challenge submissions demonstrate novel methods for applying LLMs to real-world financial NLP tasks. For example, Finance Wizard \cite{lee2024finance} fine-tuned a LLaMA3-based model for summarizing financial news, showcasing how transformer models can be adapted to sector-specific language with minimal overhead. L3iTC \cite{pontes2024l3itcfinllmchallengetask} explored quantization techniques and Low-Rank Adaptation (LoRA) to make LLMs more resource-efficient in financial text classification and summarization. In parallel, the CatMemo team \cite{cao2024catmemofinllmchallengetask} proposed a data fusion strategy to improve cross-task generalization, integrating diverse financial datasets to fine-tune LLMs more effectively.

\section{Methods}\label{sec:architecture}
\subsection{Architecture}
Our portfolio optimization framework integrates reinforcement learning (RL) and natural language processing (NLP) with a three-tier hierarchical structure. Base agents, using Stable Baselines 3 algorithms, process monthly financial metrics from YahooFinance or sentiment scores from financial news via FinBERT, proposing portfolio weights in custom RL environments with a reward function balancing ROI, volatility, and drawdown. Meta-agents, built in PyTorch, refine outputs from data-driven and NLP-based base agents, while a super-agent combines these to produce final allocations. Trained on 2003–2017 data and backtested on 2018–2024, the system outperforms benchmarks, effectively blending quantitative and qualitative insights for modern investment strategies.
\begin{figure}[H]
    \centering
    \includegraphics[width=0.8\linewidth]{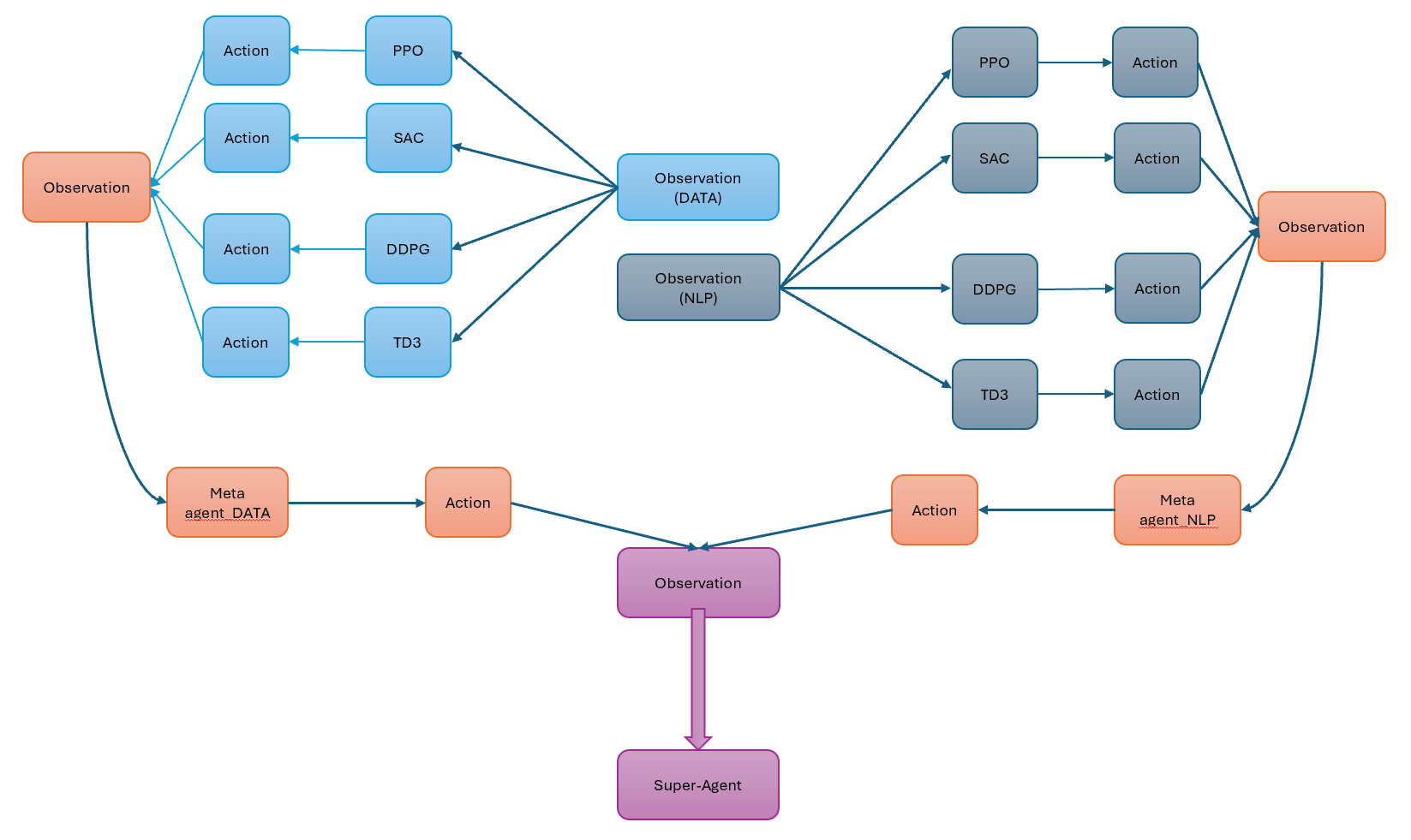}
    \caption{Summarized Architecture}
    \label{fig:architecture}
\end{figure}
\noindent\textbf{Discussion:} Why RL Instead of Classical ML Models?\\
\noindent Reinforcement Learning (RL) outperforms classical machine learning in financial portfolio optimization by excelling in sequential decision-making, adaptability, and direct objective optimization. RL agents optimize long-term rewards in dynamic markets, adapting to shifting conditions through continuous learning, unlike static ML models that struggle with non-stationary data.

\subsection{NLP-Driven Pipeline}

\subsubsection{How to Aboard the Time Specific Data Collection Issue?}
To collect news articles matching each month from 2003 to 2024, we use Google News with date filters. For each of the 14 assets, we define search terms (e.g., "S\&P 500", "SPX") and scrape articles published within each month. These articles are processed with FinBERT, a model that analyzes financial sentiment, to produce monthly sentiment scores. The pseudo code (given by algorihm \ref{alg:scrapping}) outlines this process:

\begin{algorithm}[H]
\caption{News Scraping and Sentiment Analysis}\label{alg:scrapping}
\begin{algorithmic}[1]
\State \textbf{Input:} Assets and keywords
\State \textbf{Output:} Monthly sentiment scores
\For{each asset}
    \State Define terms (e.g., "S\&P 500" = \{"SP 500", "SPX"\})
\EndFor
\For{each term}
    \For{each month in 2003–2024}
        \State Generate Google News URL with date filter
        \State Scrape the 10 first article for each links
    \EndFor
\EndFor
\For{each article}
    \State Extract text
    \State Compute sentiment with FinBERT
    \State Compute asset sentiment score \( S_t = \frac{\sum (P_{\text{positive}} - P_{\text{negative}})}{N} \)
\EndFor
\State Store scores by month and asset
\end{algorithmic}
\end{algorithm}

\subsubsection{NLP Driven Observation Vectors}
The NLP-driven observation vector for each month combines:
\begin{itemize}
    \item \textbf{Volatility Vector}: Standard deviation of daily returns.
    \item \textbf{Sentiment Score Vector}: Derived from that month's news.
\end{itemize}

We chose to stress the importance of volatility as it gives the agent an extra leg to stand on. The volatility of the market is a strong indicator and it often indicates the precision of trends (trends will be simpler to identify in a low volatility market).

\subsection{Data-Driven Pipeline}

\subsubsection{Collecting Closing Prices}
We gather daily adjusted closing prices for 14 financial asset from January 1, 2003, to December 31, 2024. This data is fetched using the \texttt{yfinance} Python library, which connects to Yahoo Finance. Adjusted closing prices are used because they adjust for events like stock splits and dividends, making them suitable for accurate financial analysis. The process involves specifying asset tickers (e.g., GSPC for S\&P 500), setting the date range, and downloading the data into a structured format like a CSV file.

\subsubsection*{Data handling}
Data quality is paramount in financial modeling, and rigorous preprocessing ensures that Reinforcement Learning (RL) agents receive clean, standardized inputs. For price data, missing values—often due to non-trading days—are addressed using forward-filling, backward filling, or linear interpolations, as financial prices typically change gradually. This method preserves the continuity of market trends by minimizing disruptions in the time series. Prices are then normalized to a 0-1 scale using min-max scaling, which is essential for comparing assets with vastly different price magnitudes. Without normalization, RL agents might inadvertently overweight higher-priced assets, skewing portfolio allocations.

\noindent For sentiment data, monthly aggregation of sentiment scores normalizes volume disparities across assets, as some indices receive far more media coverage than others. This ensures that sentiment inputs are consistent and comparable, preventing bias toward heavily covered assets.

\subsubsection{Creating Monthly Observation Vectors}
Using the daily closing prices, we create monthly observation vectors for the reinforcement learning (RL) agent. For each month, we compute:
\begin{itemize}
    \item \textbf{Sharpe Ratio}: Risk-adjusted return based on daily returns.
    \item \textbf{Sortino Ratio}: Focuses on downside risk.
    \item \textbf{Calmar Ratio}: Return relative to maximum loss.
    \item \textbf{Maximum Drawdown}: Largest drop within the month.
    \item \textbf{Volatility}: Standard deviation of daily returns.
\end{itemize}
We also compute a correlation matrix from daily returns across all assets, flattening it into a vector. These metrics form a monthly vector that informs the RL agent about market conditions.

\begin{figure} [H]
    \centering
    \includegraphics[width=1\linewidth]{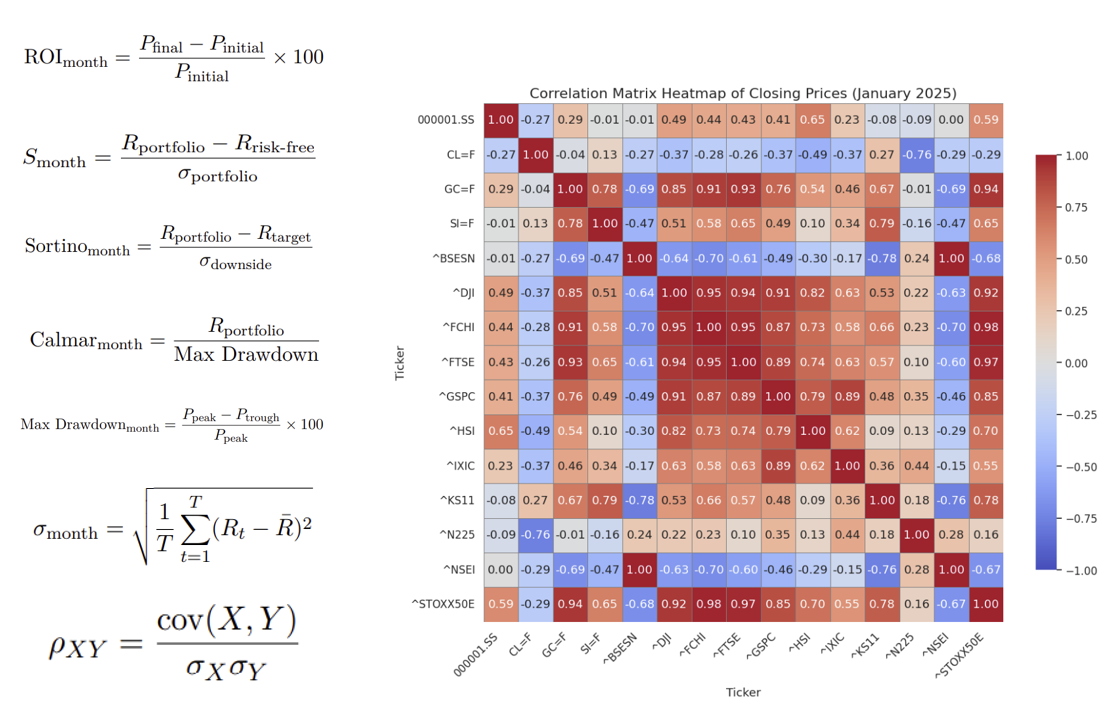}
    \caption{Metrics Computations and Correlation Matrix Example}
    \label{fig:correlation_metrics}
\end{figure}

\subsection{\small{Reproducibility and Open Access Code}}

To ensure full transparency and enable further research, all experiments presented in this paper are reproducible via three Google Colab notebooks, each addressing a different part of the pipeline:

\begin{itemize}
    \item \textbf{Data Pipeline and Sentiment Extraction:} The first notebook\footnote{\url{https://colab.research.google.com/drive/1DLQIooP7kNYHztQ7tHu5eO9NPNDPxIrY?usp=sharing}} provides a detailed, end-to-end pipeline for financial data collection and sentiment score generation. It scrapes financial news, applies FinBERT to extract sentiment at the asset level, and exports formatted sentiment scores for downstream use.
    
    \item \textbf{Fast Simulated RL Run (Sentiment Precomputed):} The second notebook\footnote{\url{https://colab.research.google.com/drive/1FPX9_8z0X39Pg3tf1bSvoByEWbbQ_juF?usp=sharing}} reproduces the reinforcement learning training pipeline using simulated sentiment data. This allows users to quickly test model dynamics, training cycles, and agent behavior with minimal compute (typically under 30 minutes).
    
    \item \textbf{Full Pipeline with Training:} The third notebook\footnote{\url{https://colab.research.google.com/drive/1SbKGmPLjF2DAKkNwEYdfc_2lS2KWMySi?usp=sharing}} combines the data scraping, sentiment extraction, and RL training into one integrated workflow. While comprehensive, this notebook is compute-intensive and requires approximately 8 hours of runtime in a typical Colab Pro environment.
\end{itemize}

This modular design offers both accessibility for quick experimentation and full reproducibility of the long-term training benchmarks presented in the paper. 

\section{Financial Instruments}\label{sec:data}

Our portfolio consists both of equities and commodities, selected to ensure diversification across asset classes, regions, and economic drivers. Stock indices capture broad market dynamics and offer lower idiosyncratic risk, while commodities reflect real-world supply-demand conditions, providing uncorrelated signals. 

\subsection{List of Assets}

To ensure sufficient market coverage and data diversity, the portfolio includes both equities and commodities spanning multiple geographic regions and economic sectors. Stock indices serve as proxies for macroeconomic conditions across developed and emerging markets, while commodities provide exposure to real asset dynamics and serve as potential hedges during equity downturns. This combination supports the training of reinforcement learning agents on heterogeneous data sources and enhances the model's ability to generalize across financial regimes.

Table~\ref{tab:all_assets} summarizes the selected instruments along with their corresponding tickers and asset class labels. These assets were chosen based on liquidity, historical availability, and relevance in global financial markets.

\begin{table}[H]
    \centering
    \resizebox{\linewidth}{!}{%
    \begin{tabular}{|l|l|l|}
        \hline
        \textbf{Ticker} & \textbf{Asset} & \textbf{Asset Class} \\
        \hline
        GSPC         & S\&P 500 Index                  & Equities \\
        IXIC         & NASDAQ Composite                & Equities \\
        DJI          & Dow Jones Industrial Average    & Equities \\
        FCHI         & CAC 40 (France)                 & Equities \\
        FTSE         & FTSE 100 (UK)                   & Equities \\
        STOXX50E     & EuroStoxx 50                    & Equities \\
        HSI          & Hang Seng Index (Hong Kong)     & Equities \\
        000001.SS    & Shanghai Composite (China)      & Equities \\
        BSESN        & BSE Sensex (India)              & Equities \\
        NSEI         & Nifty 50 (India)                & Equities \\
        KS11         & KOSPI (South Korea)             & Equities \\
        GC=F         & Gold                            & Commodities \\
        SI=F         & Silver                          & Commodities \\
        CL=F         & WTI Crude Oil Futures           & Commodities \\
        \hline
    \end{tabular}
    }
    \caption{Complete list of financial instruments used in the portfolio, grouped by asset class.}
    \label{tab:all_assets}
\end{table}

\subsection{Portfolio Management: The Basics}
Portfolio management involves strategically allocating capital across a variety of assets to optimize the trade-off between risk and return. The primary objective is to maximize returns. We try to minimize risk, often measured as the volatility or unpredictability of those returns, we implement this ideology by penalizing volatility and MDD in the rewards function for the agent.

\subsection{Portfolio Constraints and Rules}
Our experiment imposes strict rules to mimic realistic investment scenarios.
\begin{itemize}
    \item \textbf{Long-Only}: We only buy assets, not sell them short. Short-selling—borrowing an asset to sell, then repurchasing it later—adds complexity and risk (e.g. unlimited losses if prices soar). A long-only approach keeps things simpler and safer, aligning with conservative strategies.
    \item \textbf{No Leverage}: We invest only the capital we have, without borrowing. Leverage amplifies gains but also losses—borrowing \$50,000 to add to a \$100,000 portfolio could double profits or wipe out the initial stake. Avoiding leverage caps downside risk.
    \item \textbf{Monthly Rebalancing}: Every month, the RL agent reassigns weights to the 14 assets based on its policy. For example, if gold surges, it might increase gold’s share from 7\% to 10\%. This cadence balances adaptability with practicality, as frequent trading incurs costs (excluded here for simplicity).
    \item \textbf{Equal Initial Weights}: At the outset, each asset gets roughly 7.14\% of the portfolio. This neutral start lets the RL agent shape the portfolio without inherited biases.
\end{itemize}

These constraints ground the experiment in real-world norms, ensuring that AI decisions are practical and interpretable. To change those, it is possible to use the codes provided in Section 3.4 and changing or taking out parameters (for leverage, take out the normalization step) 

\subsection{Benchmarks for Performance Evaluation}
We measure our RL approach against two standards:
\begin{itemize}
    \item \textbf{Equal-Weighted Portfolio}: Each of the 14 assets gets 7.14\%, this gives an idea of the performance gains of the strategy compared to a simple buy and hold.
    \item \textbf{S\&P 500 (GSPC)}: The most commonly used financial benchmark in Portfolio Management. tracking U.S. market performance.
\end{itemize}

\begin{figure}[H]
    \centering
    \includegraphics[width=1\linewidth]{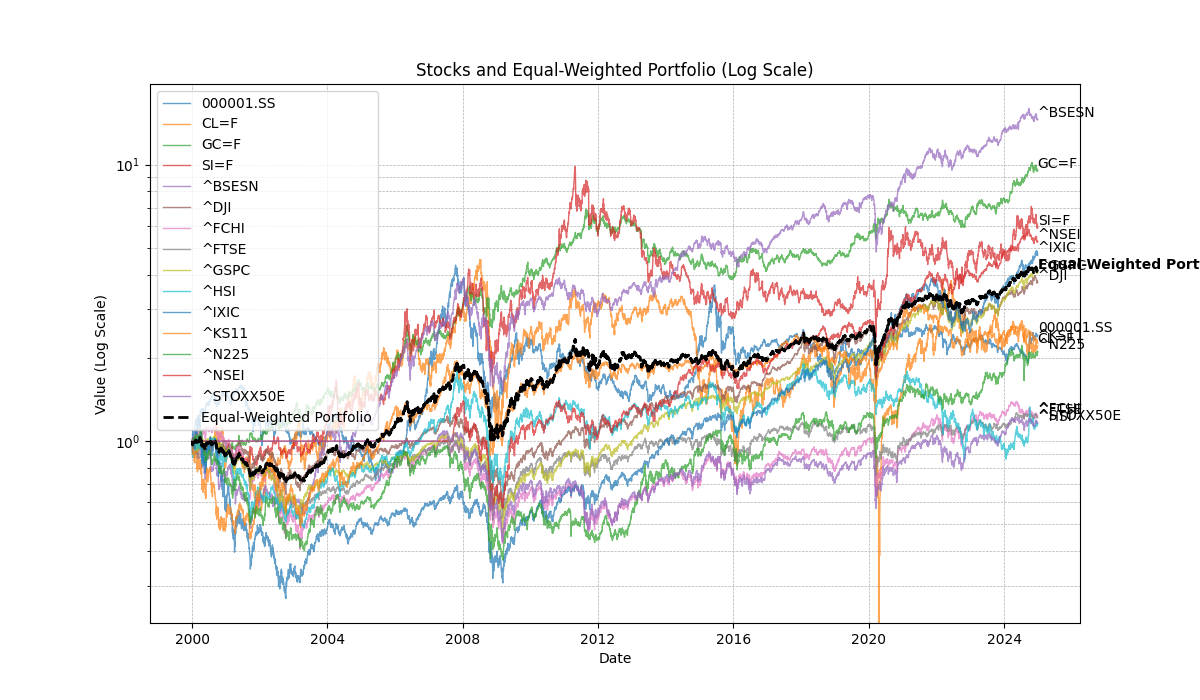}
    \caption{Log evolution of Normalized Asset Prices vs Normalized Equal weights (2003-2025)}
    \label{fig:log_evoluation}
\end{figure}

We choose to model log evolution to get a grasp of a strongly varying financial context. It would be hard to get a good idea of what is happening if using linear scales as markets change very strongly and very fast.

\section{Stable Baselines 3 Agents and Environment Setup}\label{sec:agent_description}

Stable Baselines 3 (SB3) \cite{raffin2021stable} is a widely adopted Python library that implements state-of-the-art reinforcement learning (RL) algorithms on top of OpenAI Gym environments. Its modular design, ease of integration, and support for policies make it well-suited for financial applications where agents must learn sequential allocation decisions.

In our framework, SB3 agents operate within a custom portfolio management environment. Each month, the agent observes a state vector composed of either financial indicators, sentiment signals, and outputs portfolio weights over a basket of assets. The reward function balances returns, volatility, and drawdown, allowing the agent to adaptively learn strategies aligned with financial performance objectives.

\subsection{Action}
The action space is continuous, representing the portfolio weights for each asset. These weights must sum to 1 and be non-negative (no leverage, no short-selling), aligning with standard portfolio constraints. A continuous action space allows for precise adjustments, unlike discrete actions which would limit flexibility in allocation.

\subsection{Reward Function}
The reward function guides the agent's learning by balancing multiple objectives:
\begin{itemize}
    \item \textbf{Return on Investment (ROI)}: Encourages higher portfolio returns.
    \item \textbf{Penalties}: For high volatility and large drawdowns, discouraging excessive risk.
\end{itemize}

We chose to attribute relative importance to each by a linear combination:*
\[
Reward=\alpha_1 * ROI  - \alpha_2*MDD-\alpha_3*\sigma
\]

with the $\alpha_i$'s some real values defined depending on investor needs.
For the results presented below, we used values varying between 0.5 and 2 (giving a relative but still consistent importance to each component, and severely punishing $MDD$).

\subsection{Overview of Agents}
We employ four well-established reinforcement learning algorithms tailored for continuous control in financial environments: Proximal Policy Optimization (PPO) \cite{schulman2017proximal}, Soft Actor-Critic (SAC) \cite{haarnoja2018soft}, Deep Deterministic Policy Gradient (DDPG) \cite{lillicrap2015continuous}, and Twin Delayed DDPG (TD3) \cite{fujimoto2018addressing}. PPO offers stable on-policy learning via clipped updates, while SAC encourages exploration through entropy maximization in off-policy settings. DDPG leverages deterministic policies for fine-grained action selection, and TD3 improves upon DDPG by mitigating overestimation bias with dual critics and delayed updates. (See appendix for detail)

\subsection{Backtesting}
Backtesting evaluates the RL agents on historical data to assess their performance. We test the agents on both the training period (2003–2017) and unseen data (2018–2024) to measure their ability to generalize beyond the training set.

\subsection{Seeds}
To ensure reproducibility, we use fixed consecutive seeds for all experiments. Seeds control the randomness in the environment and algorithms, allowing consistent results across runs.

\section{Hierarchy Structure}\label{sec:hierarchy_structure}

\subsection{Why Use Hierarchy Structures in AI?}
Hierarchical Reinforcement Learning (HRL) enhances portfolio optimization by decomposing decision-making into manageable components, improving interpretability, scalability, and stability \cite{sutton1999hierarchical}. Base agents specialize in quantitative financial metrics or qualitative NLP-derived sentiment scores \cite{li2021sentiment}, enabling clear, traceable decisions. Meta-agents aggregate these outputs into cohesive strategies \cite{kulkarni2016hierarchical}, ensuring transparency and ease of adjustment. This structure scales efficiently for larger portfolios or additional data types without excessive computational complexity, while meta-agents stabilize decisions by smoothing erratic actions, reducing portfolio volatility in dynamic financial markets \cite{jiang2017deep}.

\subsection{Environment Setup}

Two distinct hierarchies are established within the HRL framework. The first hierarchy consists of Natural Language Processing (NLP)-based agents, while the second is data-driven agents. By separating these tasks, the framework ensures that each base agent specializes in a specific data modality, producing traceable and interpretable recommendations.

A naive approach to combining base agent outputs might involve computing a weighted average of their recommendations for asset allocations in a given month. However, such statistical methods fail to account for the strengths and weaknesses of individual agents, limiting their ability to adapt to complex market conditions. Weighted averages or similar numerical methods lack the capacity to learn dynamically from agent performance, reducing their effectiveness in volatile financial environments. This limitation underscores the need for a more sophisticated aggregation strategy that can learn optimal policies over time \cite{jiang2017deep}.

To address this, we design a custom reinforcement learning environment implemented in PyTorch, where a meta-agent receives an observation vector formed by concatenating the proposed action vectors from each base agent across different seeds and layouts (NLP-based or data-driven). Each action vector represents a recommended weight allocation for assets in the portfolio. The meta-agent processes this observation vector and outputs a final action vector, which is a weight allocation ensuring that the portfolio. This hierarchical structure allows the meta-agent to learn how to weigh the contributions of base agents dynamically, improving decision-making in dynamic financial markets \cite{kulkarni2016hierarchical}.

Both base and meta-agents are trained on historical financial data spanning 2003 to 2017, a period that includes diverse market conditions such as the 2008 financial crisis \cite{brunnermeier2009deciphering}. The training process enables agents to learn optimal policies through interaction with the environment. The performance of the HRL framework is evaluated on a separate testing period, ensuring robustness and generalizability. \cite{jiang2017deep}.

The meta-agent is implemented as a three-layer fully connected neural network with ReLU activations and a softmax output layer, as described by the following equation:

\begin{align}
\small
f_{\theta}(X_t) = \text{Softmax}\big(& W_3 \cdot \text{ReLU} (W_2 \cdot \text{ReLU} \nonumber \\
& (W_1 \cdot X_t + b_1) + b_2) + b_3 \big)
\end{align}

where:
\begin{itemize}
    \item \(X_t\) is the observation vector at time \(t\), comprising concatenated action vectors from base agents.
    \item \(W_1, W_2, W_3\) are weight matrices for the neural network layers.
    \item \(b_1, b_2, b_3\) are bias vectors.
    \item \(\text{ReLU}(\cdot) = \max(0, \cdot)\) is the rectified linear unit activation function.
    \item The softmax layer ensures that the output portfolio weights sum to 1, satisfying the portfolio constraint.
\end{itemize}
This architecture, inspired by deep reinforcement learning frameworks \cite{mnih2015human}, enables the meta-agent to learn complex mappings from base agent outputs to optimal portfolio allocations, balancing interpretability and performance.

\section{Final Super Agent}\label{sec:super_agent}

In this section, we introduce the final super agent, which serves as the top-level decision-maker in our HRL structure. The super agent aggregates insights from lower-level meta-agents to determine the optimal portfolio allocation as explained in algorithm~\ref{alg:aggregator_train}

\begin{algorithm}[H]
\caption{Training Super-Agent using PyTorch}
\label{alg:aggregator_train}
\begin{algorithmic}[1]
\Require Trained base RL agents \( \{ A_{meta data}, A_{meta NLP} \} \), training dataset \( D_{\text{train}} \), learning rate \( \alpha \), epochs \( E \)
\Ensure Trained Meta-agent

\State Initialize PyTorch neural network \( f_{\theta} \) with random weights
\State Define loss function \( L(\theta) = \frac{1}{B} \sum_{i=1}^{B} \| f_{\theta}(X_i) - w^*_i \|^2 \)
\State Define optimizer \( \text{Adam}(\theta, \alpha) \)
\State Collect training data:

\For{each time step \( t \) in \( D_{\text{train}} \)}
    \State Compute base agent decisions \( w_t^{(i)} = A_i(X_t) \) for all agents
    \State Simulate future portfolio value for each \( w_t^{(i)} \) over \( H \) steps (lookahead reward)
    \State Select the best action \( w^*_t = \arg\max_{w_t^{(i)}} \sum_{j=t}^{t+H} R_j \)
    \State Store training sample \( (X_t, w^*_t) \)
\EndFor

\For{each epoch \( e \) in \( \{1, ..., E\} \)}
    \State Shuffle training data
    \For{each batch \( B \) in training set}
        \State Compute predictions \( \hat{w_B} = f_{\theta}(X_B) \)
        \State Compute loss \( L(\theta) \)
        \State Update model: \( \theta \leftarrow \theta - \alpha \nabla_{\theta} L(\theta) \)
    \EndFor
\EndFor

\State Return trained model \( f_{\theta} \)
\end{algorithmic}
\end{algorithm}

\subsection*{Aggregation and Observation Vectors}

The observation vector for the super agent consists of the portfolio weights proposed by the meta-agents. Specifically, it includes:
\begin{itemize}
    \item The weights suggested by the data-driven meta-agent, which focuses on quantitative metrics.
    \item The weights suggested by the NLP-based meta-agent, which incorporates sentiment analysis.
\end{itemize}
This observation vector allows the super agent to "see" the recommendations from both perspectives, enabling it to make a well-rounded decision by balancing numerical data and market sentiment.\\

\noindent We use the same structure as the meta-agents for this agent. The only changing variable is the input, which is now the concatenated action vectors of the two meta agents.
This structure strongly mimics a common way in financial markets, comparing market sentiment to current state and finding discrepancies is what gives financial actors an edge. We can see the data based meta agent as a market analyzer and the NLP based one as a conviction giver. This gives a direction from which traders can benefit.

\section{Summary of results}

Table \ref{tab:super_agent_performance} gives the reader an overview of the final results. As presented below, all meta agents beat benchmarks over the testing period and the super-agent seems to be implement a very strong strategy.

\begin{table}[h]
    \centering
    \small
    \begin{tabular}{cccc}
        \toprule
        \textbf{Agent/Benchmark} & \textbf{ROI (\%)} & \textbf{Sharpe} & \textbf{Volatility (\%)} \\
        \midrule
        Equal-Weights & 7.5 & 0.57 & 13.3 \\
        S\&P 500 & 13.2 & 0.63 & 19.7 \\
        Meta-Agent (Metrics) & 14.7 & 0.8 & 16.0 \\
        Meta-Agent (NLP) & 20.5 & 1.2 & 16.0 \\
        Super-Agent & 26.0 & 1.2 & 20.0 \\
        \bottomrule
    \end{tabular}
    \caption{Performance of super-agent vs. benchmarks and meta-agents (2018--2024).}
    \label{tab:super_agent_performance}
\end{table}

Table \ref{tab:agent_performance} provides a granular analysis of all agents, note that the results for base agents are the median out of the 5 seeds tested.

\begin{table}[h]
    \centering
    \resizebox{\linewidth}{!}{%
    \begin{tabular}{c|ccc} 
        \toprule
        \textbf{Agent} & \textbf{Annualized ROI (\%)} & \textbf{Annualized Sharpe} & \textbf{Annualized Volatility (\%)} \\ 
        \midrule
        Equal-Weights Portfolio & 7.5 & 0.57 & 13.3 \\
        S\&P 500 & 13.2 & 0.63 & 19.7 \\
        \\[-1ex]
        PPO\textsubscript{metrics} & 12.9 & 0.6 & 18.0 \\
        SAC\textsubscript{metrics} & 9.4 & 0.6 & 10.4 \\ 
        TD3\textsubscript{metrics} & 16.5 & 0.8 & 21.3 \\ 
        DDPG\textsubscript{metrics} & 10.9 & 0.5 & 18.4 \\ 
        Meta-Agent\textsubscript{metrics} & 14.7 & 0.8 & 16.0 \\
        \\[-1ex]
        PPO\textsubscript{NLP} & 14.8 & 1.0 & 13.4 \\
        SAC\textsubscript{NLP} & 9.1 & 0.9 & 10.0 \\ 
        TD3\textsubscript{NLP} & 17.5 & 0.8 & 19.2 \\ 
        DDPG\textsubscript{NLP} & 12.9 & 0.7 & 18.0 \\ 
        Meta-Agent\textsubscript{NLP} & 20.5 & 1.2 & 16.0 \\ 
        \\[-1ex]
        Super-Agent & 26.0 & 1.2 & 20.0 \\
        \bottomrule
    \end{tabular}
    }
    \caption{Analysis of Results for Agents and Benchmarks.}
    \label{tab:agent_performance}
\end{table}

\subsection*{Comparison with State-of-the-Art RL Strategies}
To contextualize our framework’s performance, Table~\ref{tab:comparison_rl_strategies} compares our meta-agents and super-agent with recent RL-based portfolio optimization strategies from academic literature. We compare our results to the 2024 study \cite{espiga2024systematic}, and against the deep RL framework \cite{jiang2017deep}. Closely competing with CNN-RL (22.0\% ROI, 1.3 Sharpe), our super agent seems to have surpassed the current state of the art. Furthermore, the consistent superiority of NLP augmented agents goes to confirm the results of \cite{xu2018news}.

\begin{table}[h]
    \resizebox{\linewidth}{!}{%
    \centering
    \small
    \begin{tabular}{lccc}
        \toprule
        \textbf{Strategy} & \textbf{Annualized ROI (\%)} & \textbf{Sharpe Ratio} & \textbf{Volatility (\%)} \\
        \midrule
        Meta-Agent (Metrics) & 14.7 & 0.8 & 16.0 \\
        Meta-Agent (NLP) & 20.5 & 1.2 & 16.0 \\
        Super-Agent & 26.0 & 1.2 & 20.0 \\
        \midrule
        DQN \cite{espiga2024systematic} & 26 & 0.8 & 38 \\
        DDPG \cite{espiga2024systematic} & 20.0 & 0.7 & 37 \\
        PPO \cite{espiga2024systematic} & 19 & 0.8 & 25 \\
        \midrule
        CNN-RL \cite{jiang2017deep} & 22.0 & 1.3 & 19.5 \\
        RNN-RL \cite{jiang2017deep} & 19.5 & 1.1 & 18.5 \\
        LSTM-RL \cite{jiang2017deep} & 21.0 & 1.2 & 19.0 \\
        \bottomrule
    \end{tabular}
    }
    \caption{Comparison of meta-agents and super-agent with state-of-the-art RL-based portfolio optimization strategies.}
    \label{tab:comparison_rl_strategies}
\end{table}

The super-agent’s ROI of 26.0\% demonstrates the effectiveness of the hierarchical approach, integrating quantitative metrics and sentiment analysis via NLP to outperform benchmarks and individual meta-agents. The strong performance of NLP-based agents, particularly TD3\textsubscript{NLP} and Meta-Agent\textsubscript{NLP}, underscores the value of sentiment-driven decision-making. 

\section{Conclusion and Future Directions}\label{sec:conclusion}
This paper introduces an innovative hierarchical reinforcement learning (RL) framework for portfolio optimization, integrating structured financial indicators with sentiment signals extracted from financial news using lightweight, domain-specific large language models (LLMs) such as FinBERT. The framework leverages a three-tier multi-agent architecture—comprising base agents that process hybrid data, meta-agents that aggregate these decisions, and a super-agent that synthesizes final portfolio allocations—enabling adaptive, interpretable, and robust decision-making in dynamic market environments.

However, the current implementation has limitations. It assumes synchronously available data inputs, which may not align with real-world asynchronous market conditions. Transaction costs are excluded, potentially overestimating practical returns, and the system has not been tested under adversarial or extreme market scenarios. Additionally, sentiment signals derived from financial news, while beneficial, may introduce noise or biases reflective of media perspectives, which could affect decision accuracy.

To overcome these shortcomings, future research will pursue several enhancements:
\begin{itemize}
    \item Asynchronous Data Integration: Incorporating real-time and asynchronous data streams to better reflect market dynamics.
    \item Transaction Cost and Stress Testing: Adding transaction cost modeling and evaluating performance under adversarial conditions to improve real-world applicability.
    \item Expanded Text Corpus: Broadening the sentiment analysis by including diverse sources such as earnings calls, regulatory filings, and social media.
    \item Larger LLMs Exploration: Comparing the efficacy of lightweight, domain-specific LLMs against larger, general-purpose models (e.g., GPT, Claude, LLaMA) to assess scalability and performance trade-offs.
    \item Possibility of strategy developments using other financial tools (End of month expiring options, Futures, Perpetuals, etc)
\end{itemize}

These advances aim to refine the robustness and generalizability of the framework, making it more suitable for practical deployment.

\newpage

\clearpage

\bibliographystyle{named}
\bibliography{main}

\end{document}